\documentclass[conference]{IEEEtran}
\IEEEoverridecommandlockouts
\usepackage{cite}
\usepackage{amsmath,amssymb,amsfonts}
\usepackage{algorithmic}
\usepackage{graphicx}
\usepackage{textcomp}
\usepackage{xcolor}
\usepackage{braket}
\usepackage[dvipsnames]{xcolor}
\usepackage{
    booktabs,
    lscape
}
\usepackage{booktabs}
\usepackage{multirow}
\usepackage{tabularx}
\usepackage{array}
\usepackage{url}
\usepackage{makecell}
\usepackage{hyperref}
\usepackage{adjustbox}
\usepackage{shellesc}
\usepackage{caption}[font=vsmall,labelfont=bf]
\usepackage{quantikz}
\newcolumntype{Y}{>{\centering\arraybackslash}X}

\def\BibTeX{{\rm B\kern-.05em{\sc i\kern-.025em b}\kern-.08em
    T\kern-.1667em\lower.7ex\hbox{E}\kern-.125emX}}
\begin{document}

\title{The Average Relative Entropy and Transpilation Depth determines the noise robustness in Variational Quantum Classifiers.\\
%{\footnotesize \textsuperscript{*}Note: Sub-titles are not captured in Xplore and should not be used}
}

\author{\IEEEauthorblockN{1\textsuperscript{st}Aakash Ravindra Shinde}
\IEEEauthorblockA{\textit{Department of Computer Science} \\
\textit{University of Helsinki}\\
Helsinki, Finland \\
shinde.aakashravindra@helsinki.fi}
\and
\IEEEauthorblockN{2\textsuperscript{nd} Arianne Meijer - van de Griend }
\IEEEauthorblockA{\textit{Department of Computer Science} \\
\textit{University of Helsinki}\\
Helsinki, Finland \\
arianne.vandegriend@helsinki.fi}
\and
\IEEEauthorblockN{3\textsuperscript{rd} Jukka K. Nurminen}
\IEEEauthorblockA{\textit{Department of Computer Science} \\
\textit{University of Helsinki}\\
Helsinki, Finland \\
jukka.k.nurminen@helsinki.fi}
}
\maketitle

\begin{abstract}
Variational Quantum Algorithms (VQAs) have been extensively researched for applications in Quantum Machine Learning (QML), Optimization, and Molecular simulations. Although designed for Noisy Intermediate-Scale Quantum (NISQ) devices, VQAs are predominantly evaluated classically due to uncertain results on noisy devices and limited resource availability. Raising concern over the reproducibility of simulated VQAs on noisy hardware. While prior studies indicate that VQAs may exhibit noise resilience in specific parameterized shallow quantum circuits, there are no definitive measures to establish what defines a shallow circuit or the optimal circuit depth for VQAs on a noisy platform. These challenges extend naturally to Variational Quantum Classification (VQC) algorithms, a subclass of VQAs for supervised learning. In this article, we propose a relative entropy-based metric to verify whether a VQC model would perform similarly on a noisy device as it does on simulations. We establish a strong correlation between the average relative entropy difference in classes, transpilation circuit depth, and their performance difference on a noisy quantum device. Our results further indicate that circuit depth alone is insufficient to characterize shallow circuits. We present empirical evidence to support these assertions across a diverse array of techniques for implementing VQC, datasets, and multiple noisy quantum devices.
\end{abstract}


\section{Introduction}
 Variational Quantum Algorithms (VQAs) have emerged as a promising approach for Noisy Intermediate-Scale Quantum (NISQ) devices for their potential noise-resilience and their adaptability to a wide range of problems \cite{cerezo2021variational}. In the context of machine learning, especially for classification tasks, it takes several forms of Variational Quantum Classifiers (VQCs). While extensive research has been conducted to enhance these algorithms, their implementation often relies on simulations in noise-free environments using classical hardware. This raises concerns about the reproducibility of results on actual quantum hardware. Although VQAs exhibit noise tolerance in shallow circuits \cite{sharma2020noise}, there is no definition of circuit shallowness, nor a systematic framework to predict when simulated performance translates to noisy hardware. Addressing this gap is essential for the reliable deployment of VQCs in real-world quantum systems.

Several challenges arise when training or testing VQAs on noisy quantum devices. Noise significantly impacts the overall implementation and skews the results for any Quantum Circuit implementation \cite{brun2019quantum}. VQAs are hybrid quantum-classical algorithms, and their training can be challenging due to significant changes in unitary operations for the optimizer \cite{cerezo2021variational}. This noise interaction can result in the noise-induced barren plateau problem, where gradients become negligible \cite{wang2021noise}.

Furthermore, running algorithms on quantum hardware while optimizing with classical methods is slow and often hindered by accessibility issues like high costs and long wait times for each iteration \cite{phalak2023shot}. Since VQAs need multiple iterations to find optimal solutions, these approaches can be economically inefficient.

Using noisy quantum hardware to validate simulated models often yields disappointing or unclear results due to model variations. Additional concerns with variational quantum algorithms (VQAs) include scalability and the barren plateau issue during training. These factors may suggest that VQAs are inconsequential in the development of quantum algorithms.

However, research has indicated that VQAs could be noise-resilient and particularly useful for NISQ (Noisy Intermediate-Scale Quantum) devices \cite{sharma2020noise}. Despite these findings, discrepancies have emerged in empirical executions on NISQ devices; some quantum circuits perform well, while others do not \cite{khanal2023evaluating}. Another advantage of VQCs is their tolerance to a certain level of uncertainty in results. While algorithms that demand extreme precision struggle in this regard, the probabilistic design of VQC models accommodates some level of uncertainty and is even welcomed in certain cases of machine learning implementation. Nonetheless, significant uncertainty in VQC models can still lead to inaccurate outcomes. 

Looking into the pros and cons of VQAs, we address the following questions in this manuscript. What makes some shallow quantum circuits display noise resilience during experimentation, while others with the same transpilation depth do not? Secondly, is it possible to verify if a classically simulated and trained quantum classification algorithm would perform equivalently in a noisy setting, and can it be known without full noisy executions? Finally, can this determination of noisy resilience verification lead to scalable VQAs?

Answering the posed questions, we have resulted in the following list of contributions from this research:
\begin{itemize}
    \item We propose a classical evaluation procedure to determine the reproducibility of Variational Quantum models on a noisy quantum device based on the average relative entropy between classes and transpilation depth.
    \item We present an analytical explanation for the noise-resilience of a certain shallow circuit under similar depth and architecture. 
    \item We conduct a detailed evaluation of the procedure through extensive experimentation, considering a variety of parameters. This includes different ansatzes, model implementations, datasets, and measurement techniques across various quantum devices from different providers. Each device consists of diverse transpilation methods, qubit mapping strategies, gate sets, noise levels, and architectural configurations, allowing us to cover multiple possible scenarios comprehensively.
\end{itemize}

This manuscript is structured as follows. Section \ref{Background} discusses former work in the fields of Variational algorithms, Relative Entropy, and noise. This section also provides a brief reasoning for the pursuit of this research. In the following, Section \ref{Exp. Comp.} we discuss the experimental components and in section \ref{Exp. pros.} we discuss the experimental process associated with those components. In Section \ref{Results} we present our findings, which are further analyzed in Section \ref{Discussion}. Finally, we present our conclusion in Section \ref{Conclusion} with continuations for future work in Section \ref{Future Work}.

\section{Background}\label{Background}
This section summarizes prior research on the noisy execution of Variational Quantum Algorithms and their inter-reliability using relative entropy measures. We highlight evidence showing that Variational Circuits can be noise robust in certain cases, alongside research indicating the negative impact of noise on these circuits. Additionally, we address findings on the deterioration of relative entropy in noise channels and our observations on using average relative entropy as a reproducibility measure.

Sharma et al. (2020) \cite{sharma2020noise} claim that VQAs, especially Full Matrix Compiling (FMC) and Fixed Input State Compiling (FISC), are resilient to measurement and depolarization noise, supported by theoretical insights and numerical results across various hardware. They attribute this resilience to a general unitary transformation, but do not explore the parameterized resilience to coherent noise, leaving training, accuracy, and overall noise resilience unexamined. Fontana et al. (2021) \cite{fontana2021evaluating} analyze short quantum circuit simulations across different noise models, noting that the impact on VQE increases with circuit depth. In contrast, Khanal et al. (2023) \cite{khanal2023evaluating} report effects of noise on various ansatz designs and dataset complexities, without clear explanations for observed discrepancies.

Ahmed et al. (2025) \cite{ahmed2025comparative} conducted similar experimental research on QNN models to analyze the effects of noise. They aimed to explain the QNN model's resilience to simulated noise through measurement probability distribution. However, our results contradict their argument that some models perform similarly regardless of the distribution. 

Alternatively, the downsides of noise on VQAs have been discussed in several articles before, which predominantly lie with the parameter shift due to noise during training, leading to noise-induced barren plateau \cite{wang2021noise}. Similarly, Barligea et al. (2025) \cite{barligea2025scalability} presented research involving VQE and QAOA algorithms under stochastic noise and raised concerns regarding the scalability of VQE algorithms under such conditions. In Cerezo et al. (2021) \cite{cerezo2021variational}, the authors mention the idea of noise accumulation and its association with depth in a VQC without a clear discerning way to determine the ideal depth for a shallow circuit. 

Error detection techniques \cite{adermann2026variational} and error mitigation techniques \cite{de2023limitations} have been researched for noisy VQA executions, but this has raised concerns regarding the quality of the ancillary system and the scalability of applied methods. Hence, a method that determines the working of a Variational Algorithm in a noisy hardware environment could significantly enhance the usability of VQC algorithms. 

An important proof presented in De Palma et al. (2023) \cite{de2023limitations} shows that the relative entropy decreases as it passes through a noisy channel. This establishes a congruence between relative entropy, noisy circuit depth, and accuracy of the final state.

Entropy is a highly versatile concept in quantum information; this is also evident in VQAs.The entropy concept is prominently used as a cross-entropy cost function in Variational Quantum Classifiers, akin to classical neural networks. This loss function evaluates model performance by measuring output probabilities and analyzing measurement probabilities in single-qubit states \cite{hur2022quantum}. Mondal et al. (2025) \cite{mondal2025relative} used relative entropy in Quantum Neural Networks to measure divergence between learned and target states, improving training stability against CPTP depolarizing noise. Tsuda et al. (2009) demonstrated that the matrix-exponentiated gradient update, inspired by quantum relative entropy, has applications in classical machine learning for establishing an upper bound on total loss for the proposed online algorithm in \cite{tsuda2009machine}.  The quantum relative entropy plays a significant role in the Quantum Boltzmann Machine, another aspect of QML, as an intuitive way to measure the quality of the approximation \cite{biamonte2017quantum}.

%Another aspect of Quantum Machine Learning, where relative entropy plays a significant role, is the Quantum Boltzmann Machine, which is a quantum probabilistic model that is based on the Boltzmann distribution of a quantum Hamiltonian \cite{amin2018quantum}. While training a Quantum Boltzmann Machine, we aim to learn a set of Hamiltonian parameters $w_j$, for a fixed set of $H_j$, such that our input state $\rho_{train}$ is well approximated by the state $\sigma = e^{-\Sigma_jw_jH_j}/Tr(e^{-\Sigma_jw_jH_j})$. For Boltzmann machines, the quantum relative entropy $\mathcal{D}(\rho_{train}||\sigma) \equiv Tr{\rho_{train}[\log(\rho_{train}) - \log\sigma]}$ is the most intuitive way to measure the quality of the approximation. Assuming the kernels of $\rho$ and $\sigma$ coincide, the quantum relative entropy provides an upper bound on the distance between the two states, minimizing this value reduces the error in approximating the state effectively \cite{biamonte2017quantum}.

Variational Quantum Circuits have been employed to estimate Von Neumann entropy, Rényi entropies, and Quantum Mutual Information. Researchers in Goldfeld et al. \cite{goldfeld2024quantum} developed a quantum neural estimation algorithm that utilizes parameterized quantum circuits and classical neural networks to efficiently approximate various quantum information measures through sampling and optimization. Concurrently, another study \cite{shin2024estimating} introduced Quantum Mutual Information Neural Estimation (QMINE), which leverages quantum neural networks to minimize a loss function for Von Neumann entropy and quantum mutual information, utilizing a quantum version of the Donsker-Varadhan representation (QDVR).

\begin{figure}[tb]
\centering
    \includegraphics[width=0.4\textwidth]{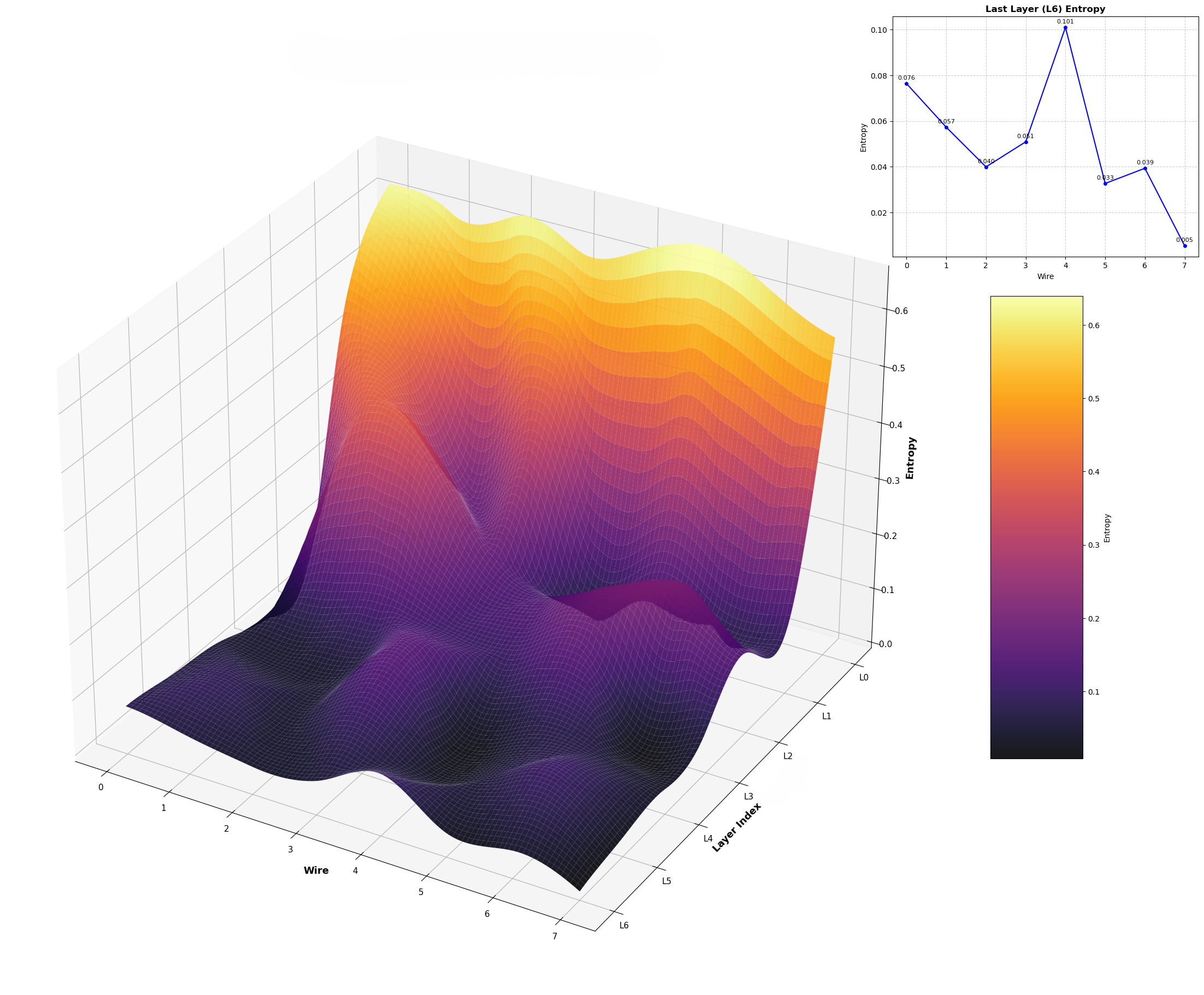}
    \caption\centering{\footnotesize Average relative entropy for data exhibiting the same measurement value across each layer and qubit, with Wire 7 being the measurement qubit. The graph on top represents the entropy values in the last layer for all qubits.}
    \label{fig:same}
\end{figure}

\begin{figure}[tb]
\centering
    \includegraphics[width=0.4\textwidth]{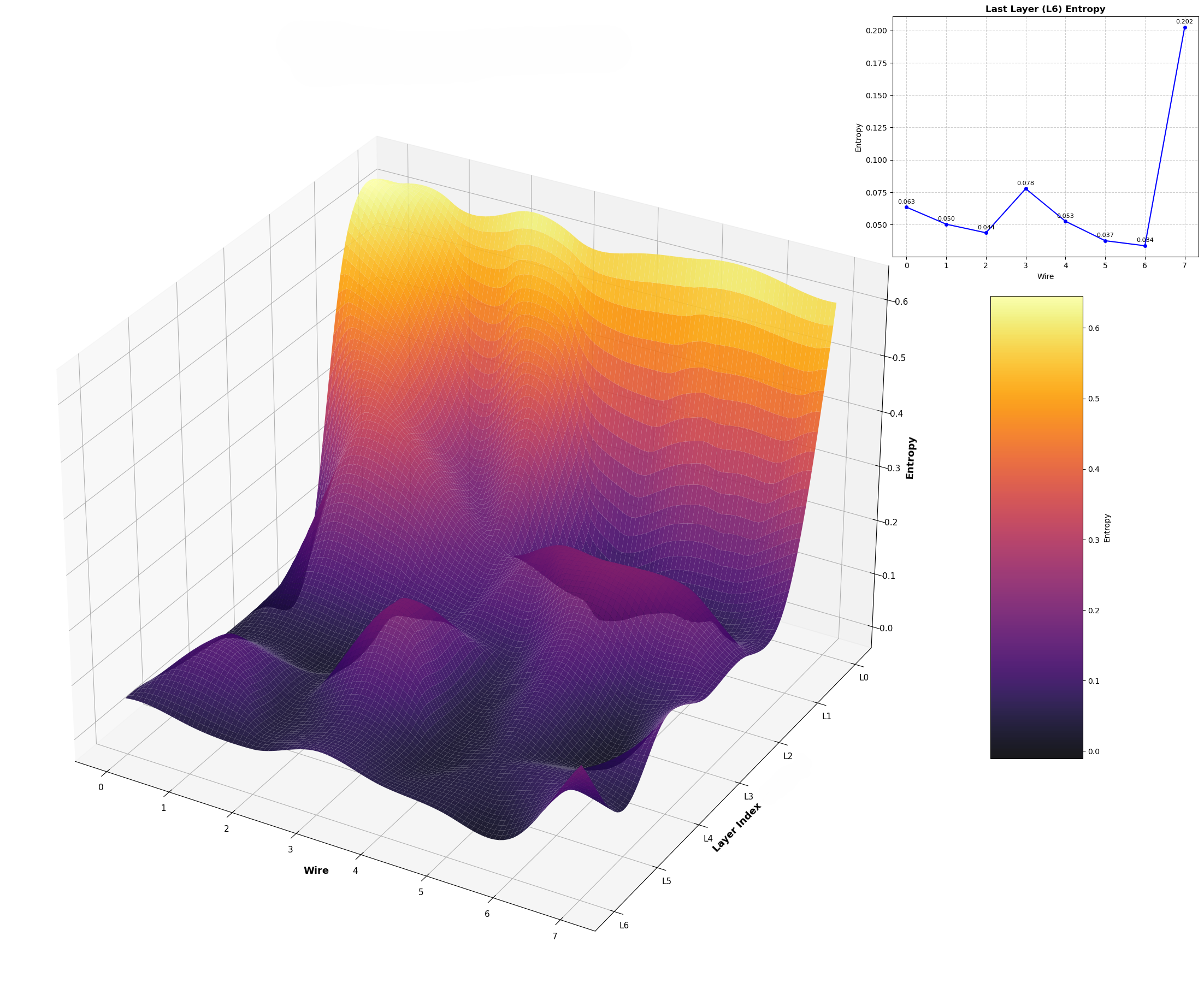}
    \caption\centering{\footnotesize Average relative entropy for data exhibiting the different measurement value across each layer and qubit, with Wire 7 being the measurement qubit. The graph on top represents the entropy values in the last layer for all qubits.}
    \label{fig:opposite}
\end{figure}

We further explored the relationship between relative entropy and variational quantum classification circuits to better understand information flow within a quantum circuit. During our initial investigation, we identified two key observations that inspired the method proposed in this article. We calculated the relative entropy for each qubit in relation to the final measurement qubit for each layer. The results are illustrated in Fig. \ref{fig:same}, which compares the average relative entropy for data exhibiting the same measurement value, and in Fig. \ref{fig:opposite}, which compares the average relative entropy for data with opposite measurement values. 

The observations from Figures Fig. \ref{fig:same} and Fig. \ref{fig:opposite} reveal that in a Variational Circuit, the entropy value decreases from the beginning to the end of a quantum circuit. Additionally, the relative entropy between opposite measurement states is significant only at the measurement qubits, as observed on Wire 7 and Layer 6 in both figures. These findings, combined with previous research by De Palma et al. (2023) on the reduction of relative entropy in a noisy channel \cite{de2023limitations}, prompted us to explore the effects of noisy executions on Variational Quantum Circuits (VQCs) by analyzing the Average Relative Entropy between classes.

%Considering the correlation between relative entropy and variational quantum circuits, we further investigated these methods in tandem. We came across interesting findings in relation to the relative entropy difference between two classification states. The first observation is that relative entropy reduces throughout the Variational Circuits as seen in Fig. \ref{fig:same} and Fig. \ref{fig:opposite}. Second observations and more important observation, as seen in Fig. \ref{fig:same}, when you calculate the relative entropy between the states that have the same measurement across individual layers, and when you calculate relative entropy between different measured states as in Fig \ref{fig:opposite}, the major difference in relative entropy values are observed at the measurement qubit. In Fig. \ref{fig:same} and Fig. \ref{fig:opposite}, the measurement qubit is the Wire 7 qubit in Layer L6, where you can clearly see the contrast in relative entropy. These initial findings regarding the association of VQC algorithms and average relative entropy (Eq. \ref{avg_ent_cal}) lead us to a deeper investigation into their relation with shallow noise robust Variational Circuits. 

In this manuscript, we expand our research for finding an association between noise, depth, and average relative entropy based on observations regarding the average relative entropy amongst classes, noise resilience in shallow depth circuits, and depleting relative entropy in a noise channel. We also intended to investigate whether this procedure could be deduced without explicitly evaluating on noisy hardware. We provide extensive empirical experimentation for building this association across various noisy backends and variational structures.  

\section{Experimental Components}\label{Exp. Comp.}
To be thorough with the empirical studies, considering the several ways in which variational classifiers could be defined, we restricted ourselves to binary classification problems with 8-qubit circuits, so that we can investigate a wide set of parameters for a generalizable result. 

Considering the different ways to represent and create variational classifiers, we evaluated three major ways in which variational circuits are implemented. Since there is no intuitive method to determine which quantum ansatzes and data encoding methods are appropriate for specific problems, we employed over 20 distinct quantum ansatz structures and 5 ways for data encoding. Additionally, we examined several other ansatz representations to identify commonalities with other quantum algorithms, as discussed in section \ref{Appendix}. As the result should not be data dependent, these models were trained on five different datasets chosen based on their complexity and variance from one another. Parameters for datasets were trained with random parameters at initialization. We used several different optimizers, such as Nestrov Momentum, Adam, Gradient Descent, and AdaGrad, with some novel variations in training as discussed in Section \ref{Exp. pros.}, as well as several different cost functions, such as mean absolute error, hinge loss, cross entropy, and square loss. Considering how all variational circuits were initialized at random, we trained all combinations at least five times. To eliminate doubts that these methods are only effective for a specific accuracy value range, we selected several models with an accuracy higher than 85\% for further evaluation.

This section offers a comprehensive overview of all the experimental components utilized in the procedure. The following section will outline how these components were implemented in the overall experiment. A detailed discussion of the experimental components is presented as follows:

\subsection{Variational Quantum Machine Learning}
Variational quantum circuits have been extensively studied in quantum machine learning. These circuits can be expressed in the form $f(\theta, x) = \mathrm{Tr}[O (x, \theta) \rho(x,\theta)]$,  where \( x \) represents the input data points, \( O \) denotes observable parameters, \( \rho \) is the density matrix, and \( \theta \) is the adjustable parameter. 

To train these models, a differentiable cost function is defined as $ \textit{L} (\theta, X, y) = \frac{1}{N}\sum_i \textit{l}(\theta, x_i, y_i)$. Here, the loss $\textit{l}(\theta, x_i, y_i)$ quantifies the model's performance for a specific training data point \( x_i \), for which the true label \( y_i \) is known. Let matrices \( X \) and \( y \) summarize all training inputs and labels, respectively \cite{bowles2024better}.

Similar to the classical neural network training procedure, the loss is numerically minimized using a classical optimizer over each iteration and batch of data values. Optimizer for each iteration updates the rotational parameters to reduce the loss value \cite{bowles2024better}.
For this experimentation, we have considered several kinds of structure, some with repeating 2-qubit ansatzes and some confined to a static structure. The 8 qubits ansatz structure was inspired by the works presented in Hubregtsen et. al. (2021) \cite{hubregtsen2021evaluation}, Hur et. al. (2022) \cite{hur2022quantum}, and Sim et. al. (2019) \cite{sim2019expressibility}. We classified these structures into three major ways of implementing the QML algorithm for this experimentation. These are classified based on an implementational naming convention as per the structures and not based on any other way of classifying methods. 

%Add appendix ref if able to draw structures
\subsubsection{Parameterized or Variational Quantum Circuits (PQC/VQC)}
Here, we define Parameterized Quantum Circuit (PQC)-based models as multi-layered ansatzes composed of both single and two-qubit quantum gates, without subsequent reduction of qubit operations. The depth of these circuits can be enhanced through the repetition of gate structures, tailored to meet the complexity requirements of the model, thus yielding improved performance outcomes. Data encoding structure is appended once at the beginning of the structure.

While increasing the repetition of the ansatzes enhances the expressibility of PQC models, it also leads to an increase in circuit depth and associated implementational constraints. This trade-off between expressibility and practicality is a critical aspect to consider in designing PQC models.  

\subsubsection{Quantum Neural Networks (QNN)}
QNNs are inherently similar to PQCs, with one key structural difference between PQCs and QNNs being that in QNNs, as depicted in  \cite{hur2022quantum}, the ansatzes acting on all qubits are halved with each layer, ultimately leading to the measurement of the remaining qubit.

QNNs have a reduced number of operational ansatzes compared to PQCs/VQCs, which helps decrease the overhead associated with circuit implementation. They have also been shown to significantly reduce the issues related to the Barren Plateau problem during the training phase. However, there are ongoing questions about the ease of simulating such structures and the expressibility of these models \cite{bermejo2024quantum}.

\subsubsection{Data Re-upload}
Data re-uploading involves repeatedly encoding classical data into quantum circuits, which can significantly improve the expressiveness of quantum models without a substantial increase in computational requirements. 

The process of data re-uploading is accomplished by linking together repeated units of tunable PQCs and Static Data encoding circuits in a sequence. Consequently, a quantum circuit can be structured as a combination of data re-uploading and single-qubit processing units. Several recent studies suggest that data re-uploading can enhance the performance and trainability of quantum models \cite{schetakis2025data}.

\subsection{Average Relative Entropy between a set of states}
Before defining Average Relative Entropy between a set of states, we first examine the definition of Quantum Relative Entropy(Von Neumann relative entropy) between a density operator $\rho \in \mathcal{D}(\mathcal{H})$ and
a positive semi-definite operator $\sigma \in \mathcal{L}(\mathcal{H})$. It is defined as,
\begin{equation}
    \mathcal{D}(\rho||\sigma) \equiv Tr{\rho[\log\rho - \log\sigma]},
\end{equation} if the $\mathrm{supp}(\rho)\subseteq \mathrm{supp}(\sigma)$ \cite{wilde2011classical}.
This measure has the same statistical interpretation as the classical relative entropy: it quantifies the distinguishability between the two quantum states $\rho$ and $\sigma$. The quantum relative entropy is an important quantity in classifying and quantifying correlations. The quantum relative entropy is 0 when $\rho = \sigma$.
%The following are the three properties of quantum relative entropy \cite{vedral2002role}:
%\begin{enumerate}
%    \item Unitary operations leave $\mathcal{D}(\rho||\sigma)$ invariant, i.e. $\mathcal{D}(U \rho U^\dagger||U\sigma U^\dagger)$. Unitary transformations represent a change of basis, and the distance between two states should not (and does not in this case) change under this. 
%    \item $\mathcal{D}(Tr_p \rho||Tr_p  \sigma) \leq \mathcal{D}(\rho||\sigma)$, where $Tr_p$ is a partial trace. Tracing over a part of the system leads to a loss of information. The less information we have about two states, the harder they are to distinguish between them, which is what this inequality says.
%    \item The relative entropy is additive $\mathcal{D}(\rho_1 \otimes \rho_2|| \sigma_1 \otimes \sigma_2) = {D}(\rho_1 || \sigma_1 ) + {D}(\rho_2||\sigma_2)$. This inequality is a consequence of the additivity of entropy itself.
%\end{enumerate}

Here, the Average Relative Entropy between two classes for a binary classification (Label 1 or Label 2) VQC model $U$ is defined as follows:

\begin{equation}
    \label{avg_ent_cal}
    \mathcal{D}_{avg}(\rho||\sigma) =  \frac{1}{n+m}\Sigma_{i,j = 0}^{n,m}\mathcal{D}(U \rho_i U^\dagger ||U\sigma_j U^\dagger  ) %\forall \{i,j\}
\end{equation}

where $i\in \{0,n\}$ for $n$ values in a set, $j \in \{0,m\}$ for $m$ values in a set, $\rho_i = \ket{\psi_i}\bra{\psi_i}$ and $ \sigma_j = \ket{\phi_j}\bra{\phi_j}$ are density matrix representation for all data values after classical to quantum embedding for their respective label class.s. From our experimentation we observe, $\mathcal{D}_{avg}(\rho||\sigma) \approx \mathcal{D}_{avg}(\sigma||\rho)$. 

\subsection{Dataset Generation Methods}\label{datasets}
For extensive evaluation, datasets with varying complexities and intricacies are essential. Hence, we employed five different methods for dataset generation, each with its own prediction difficulties. For each data generation, we kept the modifiable parameters constant, maintaining significant complexities. The first two of the dataset generators are chosen for their ability to generate Gaussian and non-Gaussian distributions, respectively. While the final three are inspired by the benchmarking initiative in \cite{bowles2024better}.
%Make Classification is a dataset generation method based on a Gaussian distribution for classifying clusters. Contrary to Make Classification, we utilize a Synthetic Dataset by Capital One for a non-Gaussian distribution for dataset creation. For the latter three datasets, we draw inspiration from et. al. \cite{bowles2024better} for selecting our datasets used for benchmarking. Details of the dataset generator methods and associated classification complexities are as follows.

\subsubsection{Make Classification by Scikit-Learn (Linear Dataset Creator)}\label{make-class}
The algorithm for the Make Classification function is based on the concept in \cite{Guyon2003DesignOE} and was developed to create the "Madelon" dataset. This function is particularly effective at introducing noise through attributes like correlated, redundant, and uninformative features. Additionally, it generates noise by forming multiple Gaussian clusters within each class and applying linear transformations to the feature space \cite{scikitlearn83Generated}.

\subsubsection{Synthetic Data by Capital One (Non-linear Dataset)}\label{synth}
The Synthetic Data generator is a tool for creating tabular datasets that incorporate non-linearity using a Non-Gaussian distribution. The feature set generated is tunable over three parameters: informative, redundant, and nuisance. Informative parameters generate features crucial for determining binary labels, which are defined using Copula theory, a mathematical approach that separates the correlation structure from the marginal distributions of the feature vectors. The Redundant parameter helps create a linear combination of the informative features, while nuisance features help generate features consisting of uncorrelated vectors drawn from the range [-1,1]. For more details about the dataset, please refer to \cite{Barr2020TowardsGT}.

\subsubsection{Hidden Manifold}\label{hid-mani}
Goldt et al. \cite{Goldt2019ModelingTI} introduced a data generation procedure designed to explore the impact of data structure, such as the size of a hidden manifold believed to influence problem difficulty, on learning outcomes. The process involves generating inputs on a low-dimensional manifold, which are then labeled by a simple, randomly initialized neural network. Finally, these inputs are projected into a d-dimensional space \cite{bowles2024better}.

\subsubsection{Hyperplanes}\label{hype}
This data generation method creates several hyperplanes in a k-dimensional space and assigns labels to randomly sampled points based on the parity of perceptron classifiers using these hyperplanes as decision boundaries. A label indicates whether a point is on the positive side of an even number of hyperplanes, aiming to reflect the "global" structure of the problem related to the arrangement of hyperplanes \cite{bowles2024better}.

\subsubsection{Two Curves}\label{two-curve}
This data generation procedure is inspired by a theoretical study, as discussed in et al. \cite{buchanan2020deep}, which demonstrates how the performance of neural networks is affected by the curvature and the distance between two one-dimensional curves embedded in a higher-dimensional space. Following the implementation in \cite{bowles2024better}, we also used low-degree Fourier series to embed two sets of data sampled from a one-dimensional interval, with one set for each class, representing these data as curves in d-dimensional space, while also adding some Gaussian noise.

\subsection{Data Encoding}
Data encoding is a key element of Quantum Machine Learning (QML). Unlike classical methods that use vectors, QML requires transforming classical data into quantum states through encoding circuits. These circuits are referred to as quantum feature maps or quantum encodings, also interchangeably known as embeddings. The encoding process significantly affects QML model performance, so choosing the right encoding method for the dataset is essential \cite{zang2025benchmarking}. We have utilized the most commonly used encoding methods for Variational Quantum Classification models in our work.

\subsubsection{Angle Encoding}
One of the most common encoding methods involves representing data values for each feature as an angle of rotation along different axes: X, Y, or Z. This is expressed mathematically as \( R_{j}(\theta) = e^{-i \frac{\theta}{2} \sigma_j} \), where \( j \in \{x, y, z\} \) and \( \sigma_j \) represents the Pauli matrices \( \sigma_x, \sigma_y, \sigma_z \). In this encoding method, N feature values are embedded into N qubits \cite{monnet2024understanding}. For our experimentation, we have tested for all three rotational values across all variational circuits in this manuscript.

\subsubsection{Amplitude Encoding}
In the case of Amplitude encoding, classical feature vectors are represented as amplitudes. Here, classical feature vector $x = \{ x_1, ..,x_n\}$ can be written as $\ket {\psi} = \Sigma_{i=1}^n x_i \ket {i}$, where $\ket {\psi}$ is the resulting quantum state and $\ket {i}$ are the quantum state for each vector components for $i$ being the binary index of values. An N-dimensional amplitude encoded classical vector requires $\log_2{N}$ qubits. Considering the expensive circuit implementation of state initialization, evaluation of amplitude encoding was essential in understanding the bottleneck of QML implementation \cite{monnet2024understanding}.

\subsubsection{Instantaneous Quantum Polynomial (IQP) Encoding}
The instantaneous Quantum Polynomial-time (IQP) encoding maps features into qubits via diagonal gates in an IQP circuit. An IQP circuit is a type of quantum circuit that consists of a sequence of Hadamard gates followed by a sequence of diagonal gates in the computational basis. In this context, the diagonal gates are single-qubit Rz rotations applied to each qubit, encoding the \(n\) features. This is followed by two-qubit ZZ entanglers, represented as \(e^{-i x_i x_j \sigma_z \otimes \sigma_z}\). This method is capable of encoding N data points into N qubits \cite{havlivcek2019supervised}. 

\subsection{Measurements}
Measurement in a quantum system involves retrieving information~\cite{11134353}. In the context of a classification Quantum Machine Learning (QML) system, the focus is on measuring quantum circuits to extract information based on quantum-encoded classical data. This process helps us determine the class for a specific set of labels. 

For binary classification problems, measurements usually involve obtaining information from a single qubit. In the practical implementation of Variational QML algorithms, measurements are often performed probabilistically, where the classes are represented by the labels '0' and '1' \cite{phalak2023shot}. Alternatively, measurements can be made as the expectation value of the system, with the labels corresponding to the eigenvalues '-1' and '1' \cite{seo2026multivariate}.

\subsubsection{Shot-based probabilistic measurement}
A "shot" in a quantum circuit is a single execution that produces probabilistic measurement results of 0 or 1 for each qubit. In Variational Circuits, quantum information is projected into the computational basis, and measurement probabilities are estimated over multiple shots, as several executions are needed for accurate probability estimation.

Although increasing the number of shots improves precision, it also raises resource costs. The statistical error in probability estimation decreases with the square root of the number of shots used. Typically, a default value of 1,024 shots is used, but this can be adjusted according to the desired level of precision.

\subsubsection{Expectation Values}
The expectation value is the probabilistic average of the results (or measurements) from an experiment. It represents an average of all possible outcomes, weighted by their likelihood. However, it is important to note that the expectation value is not necessarily the most probable value of a measurement; in fact, it can even have zero probability.

Given an observable $A = \Sigma_k \lambda_k \ket{k}\bra{k}$, we obtain eigenvalue $\lambda_k$ with probability $p_k$, we
can naturally define an expectation value of this observable in the state $ \ket{k}$ as $\braket{A_\psi} \equiv \Sigma_k \lambda_k \bra{\psi}P_k\ket{\psi} =\bra{\psi} (\Sigma_k \lambda_k P_k) \ket{\psi} = \bra{\psi}A\ket{\psi} = Tr(A\ket{\psi}\bra{\psi})$ \cite{brun2019quantum}. In our case, as measured in the Z basis, each time we will get $+1$ or $-1$. 

\section{Experimental Procedure}\label{Exp. pros.}
Experiments were conducted in three phases. In the first phase, models were trained on the datasets generated using methods mentioned in Section \ref{datasets} under ideal simulation conditions (without noise), and models that achieved an accuracy greater than 85\% were selected for the second phase. During the second phase, these models were evaluated on noisy quantum backends using the same testing data as in the previous phase. The third phase involved data analysis, where we examined several factors that could explain the differences in accuracy among various noisy quantum backends for shallow quantum circuits with similar transpilation depths. With extensive analysis, we concluded that the Average Relative Entropy between classes (as computed in Eq. \ref{avg_ent_cal}) of the quantum circuits best explains the phenomenon.

Phase one of the implementation was carried out entirely using the Pennylane quantum programming framework \cite{pennylane2018arXiv181104968B}. We trained five models for each combination of dataset, ansatz, and data encoding method mentioned in Section \ref{Exp. Comp.}. As parameter initialization is random, this approach allowed us to explore which datasets were best suited to specific quantum circuit structures.

We trained more than 2,750 models for the VQC-QNN circuits using shot-biased probabilistic measurement and another 2,750 models for expectation-value measurements. Additionally, we trained 3,000 models for data-reupload circuits with both types of measurements. Each dataset generator produced 5,000 normalized data points, which were then divided into an 80-20 train-test split. During the initial training phase, we observed that using multiple optimizers with varying batch sizes and learning rates in tandem significantly reduced the number of training iterations while achieving respectable accuracy more quickly than using a single optimizer with a fixed batch size and learning rate. Further research on this phenomenon is underway, as discussed in Section \ref{Future Work}.

From the QML models, we selected only those models that achieved an accuracy of over 85\% on the test dataset for phase two. This selection resulted in 413 models for the VQC-QNN circuits with probabilistic measurement, 362 models for expectation value measurements, and 400 models for the data-reupload circuits with both measurements. The Puhti Supercomputer was used for training these QML models in a simulated environment, considering the volume of data generation and processing involved.

In phase two, we tested QML models on noisy quantum devices using the same test dataset. Most runs were performed on simulated-fake quantum backends, due to the high costs associated with running experiments on real quantum devices and the extent of data needed for evaluation. These simulations were carried out on the Puhti Supercomputer.

To validate our results, we further tested them on actual quantum devices under several constraints. We employed multiple noisy devices from different vendors to strengthen our findings. 

From IBM, we used simulated quantum devices such as 'Sherbrooke,' 'Prague,' 'SydneyV2,' 'JohannesburgV2,' 'MelbourneV2,' 'BoeblingenV2,' 'GuadalupeV2,' 'AlmadenV2,' 'SingaporeV2,' 'Marrakesh,' 'Fez,' and 'Torino.' We confirmed our results with available IBM real quantum devices, specifically Marrakesh, Fez, and Torino, which are currently accessible on the IBM cloud. 

From IQM, we utilized simulated quantum devices like Apollo and Aphrodite, both of which have noise models and architectures similar to the real quantum devices Garnet and Emerald, respectively. Additionally, Aphrodite bears resemblance to the Q50 quantum device available at VTT, which was also used for verification. 

From IonQ, we worked with the Aria 1 Noisy Quantum Simulator, but we were unable to verify it against the actual Aria 1 device due to restricted access.

Several changes were necessary when executing Pennylane-based circuits on IQM and IonQ devices due to compatibility issues. Hence, we converted the code from Pennylane's compiled tape method into QASM circuits \cite{openqasmOpenQASMLive}. Subsequently, these QASM circuits were transformed into Qiskit \cite{qiskit2024} circuits for execution on the respective devices.

In phase three, the data obtained from these noisy backends are evaluated using transpiled depth and average relative entropy between classes. Transpiled depth values are obtained after the quantum circuit is converted into the gate set of the quantum device, using the depth function from the Qiskit implementation. Average relative entropy calculations are computed using the test datapoints for the independent classes.

Results obtained from the analysis in phase three are provided in the next section. We have also provided a graph that showcases the correlation between accuracy difference, average relative entropy, and transpilation depth for different quantum devices.

\section{Results}\label{Results}

Table \ref{tab:results} summarizes the results from our experiments, showcasing 51 distinct data points that represent similar observations across different models, encodings, and ansatzes. The results in the table consist of the relevant quantum devices for which the device provider has a noise model to verify. Table \ref{tab:results} shows values for individual models. 'Sim. Acc.' indicates the simulated accuracy, 'Ansatz' refers to the circuit structures from Section \ref{Appendix}, 'Avg Rel Ent.' is the average relative entropy between binary classes as explained in Eq. \ref{avg_ent_cal}, 'Encoding' denotes data encoding methods, and 'Model' describes circuit structures and measurement types from Section \ref{Exp. pros.}. The remaining columns provide details for test set accuracy (Acc.) and transpilation depth (Depth) of circuits on the noisy quantum backend.

Values shown in green indicate that the accuracy of the noisy device differs from the accuracy of the simulation by less than $0.03$. Values in red represent an accuracy difference greater than $0.08$, while values with accuracy differences between $0.03$ and $0.08$ are shown in blue. The value difference over multiple runs for the same test data and the same device does not differ more than $\pm 0.001$ in Table \ref{tab:results}. %The transpilation depth mentioned in the table is obtained from the values provided by the vendors of the respective noisy backend in their native gate set. 

We had explored several methods, including quantum circuit complexities, classical to quantum data encoding, relative entropy distribution in quantum circuits across different layers, dataset complexities, the probability distribution of measured values, mutual information between measured and unmeasured qubits, Quantum Rényi divergence, Noisy device attributes such as T1, T2, readout error values, and the characteristics of noisy device gate sets. 

\begin{figure}[tb]
  \centering
  \includegraphics[width=0.5\textwidth]{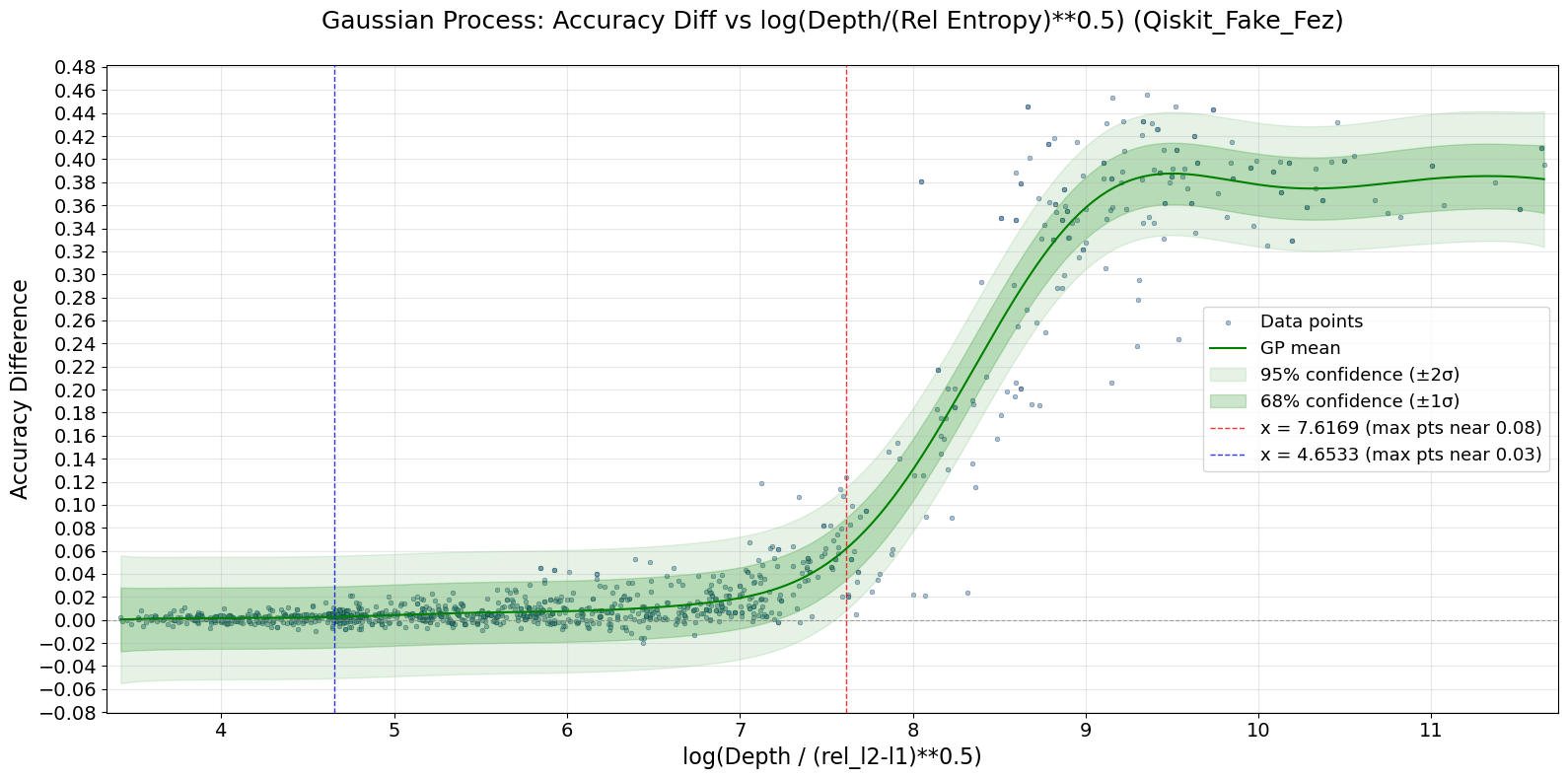}
  \caption{\centering \footnotesize Average Entropy, Transpilation Depth, and Accuracy difference relation representation using log-DTSAE for Qiskit Fake Fez Backend}
  \label{fig:graph1}
\end{figure}

\begin{figure}[tb]
  \centering
  \includegraphics[width=0.5\textwidth]{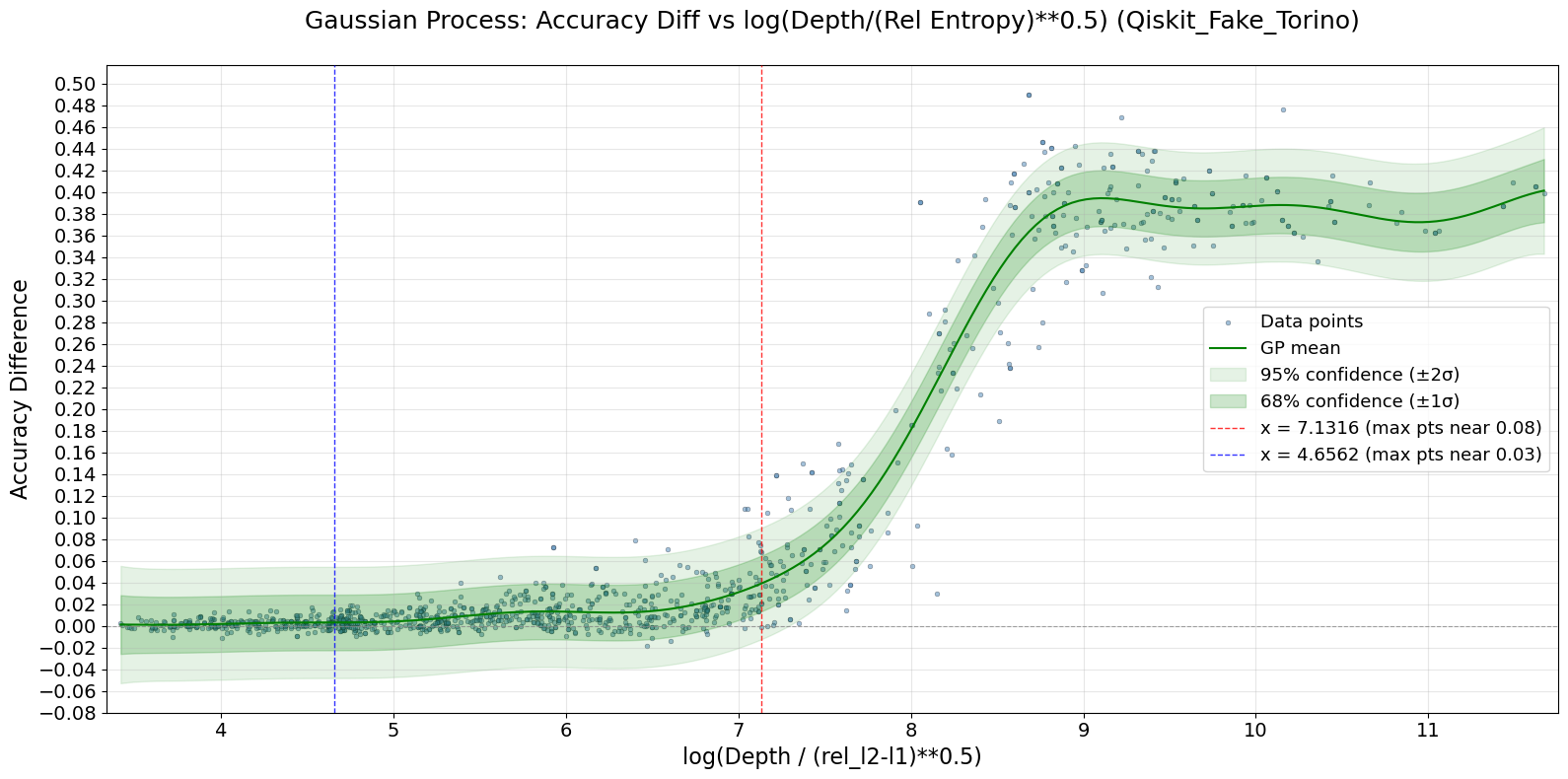}
  \caption{\centering \footnotesize Average Entropy, Transpilation Depth, and Accuracy difference relation representation using log-DTSAE for Qiskit Fake Torino Backend} 
  \label{fig:graph2}
\end{figure}

Among all these methods, we found a strong correlation between the accuracy differences observed between simulations and noisy devices, the Average Relative Entropy between classes (as computed in Eq. \ref{avg_ent_cal}), and the transpilation depth of quantum circuits before implementation on the respective noisy quantum hardware. From Table \ref{tab:results}, it is clear that there is no clear correlation between accuracy and either depth or average relative entropy by themselves as well. However, upon further evaluation, we find a correlation between the accuracy difference vs. depth and the average relative entropy. This relationship is better illustrated with the graphs in Fig. \ref{fig:graph1} and Fig. \ref{fig:graph2}.

Figures, Fig. \ref{fig:graph1} and Fig. \ref{fig:graph2} show plots of accuracy versus log-DTSAE (log of depth to square-root of average relative entropy), defined as \( \log\left(\frac{\text{depth}}{\sqrt{\text{avg\_rel\_ent}}}\right) \), based on over 1,100 analyzed models. The trends are similar across all backends, though the specific log-DTSAE values differ. The blue line indicates the point with the highest instance count, showing an accuracy difference of $0.03$, while the red line indicates a difference of $0.08$. The green GP (Gaussian Process) mean line represents a predicted growth in the accuracy difference concerning log-DTSAE. It uses Gaussian Process Regression (GPR), a non-parametric machine learning technique effective for complex continuous data \cite{wang2023intuitive}.

As seen in Fig. \ref{fig:graph1} and Fig. \ref{fig:graph2}, for a $0.03$ accuracy difference, the log-DTSAE values are $4.6533$ for Fake Fez and $4.6562$ for Fake Torino, with similar values for other backends: Fake Prague ($4.6466$), Fake Marrakesh ($4.7875$), Fake Sherbrooke ($4.719$), Fake Apollo ($3.8372$), and Fake Aphrodite ($3.8264$). For the $0.08$ accuracy difference, divergence occurs at $7.6169$ for Fake Fez, with similar progression for other backends: Fake Marrakesh($7.4310$), Fake Torino($7.1316$), Fake Prague ($7.6112$), Fake Sherbrooke($6.2366$), Fake Apollo($4.4981$), and Fake Aphrodite($4.4823$).

Due to the large amount of data required to investigate the factors influencing the decreasing accuracy on noisy backends, it was neither financially nor technically feasible to conduct the full experiment on actual quantum devices. As a result, the use of simulated backends in the experiment was necessary, which was further supported by limited experiments conducted on real quantum devices. We limited our executions to a few models with a smaller verification dataset. We compared results for accessible devices with noisy backends, including Fez(Heron R2), Torino(Heron R1), Marrakesh(Heron R2) from IBM, and Garnet(20 qubits) and Emerald(54 qubits) from IQM, with Emerald sharing architecture with Q50 hosted by VTT. 

We had the following constraints when choosing models and data points. Models were selected with varying average relative entropy and similar depth to analyze differences in the previously observed effects. We chose datapoints based on the K-Medoids to find k datapoints that best represent the overall dataset to get an overview of the differences in results across different devices. Results showed that the average variance in readout value for the selected datapoints is under $\Delta 0.03$, proving that the values associated with the noise modeled backends could be observed on the real backends as well. Extending our observations and results to the real noisy devices. More evaluations are in progress for extensive verification. All observations are available for review in the GitHub repositories mentioned in Section \ref{data}.

\begin{landscape}
\begin{table}[htbp]

\centering
\scriptsize
\setlength{\tabcolsep}{3pt}  % slightly tighter spacing

\begin{tabularx}{\linewidth}{c l l l c c *{8}{r r}}
\toprule

\multirow{2}{*}{Index} &
\multirow{2}{*}{Ansatz} &
\multirow{2}{*}{Model} &
\multirow{2}{*}{Encoding} &
\multirow{2}{*}{Avg Rel Ent.} &
\multirow{2}{*}{Sim Acc.} &

\multicolumn{2}{c}{\makecell{Qiskit \\ Fake Fez}} &
\multicolumn{2}{c}{\makecell{Qiskit \\ Fake Marrakesh}} &
\multicolumn{2}{c}{\makecell{Qiskit \\ Fake Torino}} &
\multicolumn{2}{c}{\makecell{Qiskit \\ Fake Prague}} &
\multicolumn{2}{c}{\makecell{Qiskit \\ Fake Sherbrooke}} &
\multicolumn{2}{c}{\makecell{IQM \\ Fake Aphrodite}} &
\multicolumn{2}{c}{\makecell{IQM \\ Fake Apollo}} &
\multicolumn{2}{c}{\makecell{IonQ \\ Fake Aria1}} \\

\cmidrule(lr){7-22}

& & & & & 
& Acc. & Depth
& Acc. & Depth
& Acc. & Depth
& Acc. & Depth
& Acc. & Depth
& Acc. & Depth
& Acc. & Depth
& Acc. & Depth \\

\midrule

1 & QNN\_model\_170\_pool & VQC\_QNN\_EXP & AngleY & 0.5352 & 0.892 %3jR6N
& \textcolor{ForestGreen}{0.883} & 62
& \textcolor{ForestGreen}{0.899} & 62
& \textcolor{ForestGreen}{0.901} & 62
& \textcolor{ForestGreen}{0.896}& 62
& \textcolor{ForestGreen}{0.900}& 66
& \textcolor{ForestGreen}{0.905} & 99
& \textcolor{ForestGreen}{0.895} & 99
& \textcolor{ForestGreen}{0.898} & 131 \\

2 & QNN\_H\_CZ\_RX\_pool & VQC\_QNN\_PROB & Amplitude & 0.0881 & 0.932 %QzDKVg
& \textcolor{red}{0.583} & 1472
& \textcolor{red}{0.692} & 1604
& \textcolor{red}{0.515} & 1606
& \textcolor{red}{0.883} & 1612
& \textcolor{red}{0.883} & 1804
& \textcolor{red}{0.883} & 616
& \textcolor{red}{0.883} & 578
& \textcolor{red}{0.883} & 527 \\

3 & QNN\_H\_CZ\_RX\_pool & VQC\_QNN\_PROB & AngleX & 1.5808 & 0.933 % dBR8zy
& \textcolor{ForestGreen}{0.932} & 46
& \textcolor{ForestGreen}{0.931} & 46
& \textcolor{ForestGreen}{0.935} & 46
& \textcolor{ForestGreen}{0.932} & 46
& \textcolor{ForestGreen}{0.938} & 50
& \textcolor{ForestGreen}{0.934} & 18   
& \textcolor{ForestGreen}{0.930} & 18
& \textcolor{ForestGreen}{0.934} &  25 \\

4 & QNN\_model\_170\_pool & VQC\_QNN\_EXP & IQP & 0.2337 & 0.931 % kfRoRd
& \textcolor{ForestGreen}{0.928} & 308
& \textcolor{ForestGreen}{0.918} & 320
& \textcolor{ForestGreen}{0.923} & 275
& \textcolor{ForestGreen}{0.932}& 310
& \textcolor{ForestGreen}{0.913}& 402
& \textcolor{red}{0.838} & 145
& \textcolor{red}{0.825} & 140
& \textcolor{ForestGreen}{0.924} & 78 \\

5 & H\_CZ\_RX & VQC\_QNN\_EXP & AngleY & 0.0245 & 0.929 %6UtipgwT
& \textcolor{blue}{0.884} & 54
& \textcolor{ForestGreen}{0.908} & 55 
& \textcolor{blue}{0.890} & 56
& \textcolor{blue}{0.890} & 54
& \textcolor{red}{0.843} & 144
& \textcolor{red}{0.674} & 37
& \textcolor{red}{0.656} & 37
& \textcolor{red}{0.882} & 37\\

6 & QNN\_model\_170\_pool & VQC\_QNN\_PROB & IQP & 0.6240 & 0.888 %2iQc2jSM
& \textcolor{ForestGreen}{0.89} & 292
& \textcolor{ForestGreen}{0.889} & 286
& \textcolor{ForestGreen}{0.893} & 293
& \textcolor{ForestGreen}{0.887} & 277 
& \textcolor{ForestGreen}{0.888} & 399
& \textcolor{blue}{0.844} & 127
& \textcolor{blue}{0.835} & 140
& \textcolor{blue}{0.885} & 78 \\

7 & RY\_RZ\_CNOT & VQC\_QNN\_EXP & IQP & 0.0355 & 0.897 %PZeyE2P
& \textcolor{red}{0.773} & 382
& \textcolor{red}{0.767} & 389
& \textcolor{red}{0.729} & 368
& \textcolor{red}{0.817} & 393
& \textcolor{red}{0.701} & 580
& \textcolor{red}{0.555} & 141
& \textcolor{red}{0.551} & 147
& \textcolor{red}{0.816} & 58\\

8 & RY\_CNOT & VQC\_QNN\_EXP & IQP & 0.0191 & 0.939 % duborJ9n
& \textcolor{red}{0.785} & 382
& \textcolor{red}{0.855} & 381
& \textcolor{red}{0.742} & 376
& \textcolor{red}{0.831} & 382
& \textcolor{red}{0.598} & 505
& \textcolor{red}{0.522} & 159
& \textcolor{red}{0.514} & 160
& \textcolor{red}{0.736} & 72\\

9 & RXRZ\_CNOT & VQC\_QNN\_PROB & AngleX & 0.1864 & 0.947 % Px8zEyaF
& \textcolor{ForestGreen}{0.933} & 67
& \textcolor{ForestGreen}{0.936} & 67
& \textcolor{ForestGreen}{0.934} & 67
& \textcolor{ForestGreen}{0.931} & 67
& \textcolor{ForestGreen}{0.930} & 90
& \textcolor{blue}{0.877} & 36
& \textcolor{blue}{0.872} & 36
& \textcolor{ForestGreen}{0.942} & 24 \\

10 & INV\_QFT\_U3\_MODEL & VQC\_QNN\_PROB & AngleY & 0.2127 & 0.949 % Jpk7FGY7
& \textcolor{ForestGreen}{0.947} & 147
& \textcolor{ForestGreen}{0.945} & 132
& \textcolor{ForestGreen}{0.943} & 151
& \textcolor{ForestGreen}{0.940} & 132
& \textcolor{ForestGreen}{0.937} & 181
& \textcolor{ForestGreen}{0.939} & 96
& \textcolor{ForestGreen}{0.922} & 94
& \textcolor{ForestGreen}{0.944} & 22\\

11 & RY\_CZ & VQC\_QNN\_EXP & Amplitude & 0.1710 & 0.905 % FPsQVvq
& \textcolor{red}{0.714} & 1745
& \textcolor{red}{0.744} & 1974
& \textcolor{red}{0.593} & 1981
& \textcolor{red}{0.799} & 1835
& \textcolor{red}{0.545} & 2406
& \textcolor{red}{0.531} & 642
& \textcolor{red}{0.499} & 714
& \textcolor{red}{0.787} & 594\\

12 & INV\_QFT\_U3\_MODEL & VQC\_QNN\_EXP & IQP & 0.0956 & 0.904 % GjD68jKAX 
& \textcolor{blue}{0.861} & 350
& \textcolor{blue}{0.871} & 340
& \textcolor{blue}{0.862} & 384
& \textcolor{blue}{0.862} & 385
& \textcolor{red}{0.787} & 476
& \textcolor{red}{0.587} & 197
& \textcolor{red}{0.567} & 199
& \textcolor{blue}{0.853} & 58\\

13 & QNN\_RY\_CRX\_pool & VQC\_QNN\_EXP & AngleX & 1.0540 & 0.853 % 5zqHjiD6Y
& \textcolor{ForestGreen}{0.857} & 58
& \textcolor{ForestGreen}{0.852} & 58
& \textcolor{ForestGreen}{0.860} & 58
& \textcolor{ForestGreen}{0.854} & 58
& \textcolor{ForestGreen}{0.856} & 62
& \textcolor{ForestGreen}{0.850} & 24
& \textcolor{ForestGreen}{0.854} & 24
& \textcolor{ForestGreen}{0.851} & 29 \\

14 & QNN\_model\_150params & VQC\_QNN\_PROB & AngleX & 0.6778 & 0.997 % d84TKBz
& \textcolor{ForestGreen}{0.995} & 62 
& \textcolor{ForestGreen}{0.997} & 62
& \textcolor{ForestGreen}{0.997} & 62
& \textcolor{ForestGreen}{0.999} & 62
& \textcolor{ForestGreen}{0.997} & 66
& \textcolor{ForestGreen}{0.991} & 26
& \textcolor{ForestGreen}{0.995} & 26
& \textcolor{ForestGreen}{0.995} & 26 \\

15 & QNN\_H\_CZ\_RX\_pool & VQC\_QNN\_EXP & AngleY & 0.2083 & 0.942 % 97pBDmmB
& \textcolor{ForestGreen}{0.936} & 47
& \textcolor{ForestGreen}{0.938} & 47
& \textcolor{ForestGreen}{0.931} & 47
& \textcolor{ForestGreen}{0.943} & 47
& \textcolor{ForestGreen}{0.927} & 50
& \textcolor{ForestGreen}{0.925} & 18
& \textcolor{ForestGreen}{0.916} & 18
& \textcolor{ForestGreen}{0.940} & 25 \\

16 & RXRZ\_CNOT & DATA\_REUP\_PROB & AngleX & 0.1032 & 0.925 % BnG7BaEpd4WDVBstpm3REq
& \textcolor{ForestGreen}{0.927} & 67
& \textcolor{ForestGreen}{0.919} & 67
& \textcolor{ForestGreen}{0.915} & 67
& \textcolor{ForestGreen}{0.919} & 67
& \textcolor{ForestGreen}{0.907} & 88
& \textcolor{blue}{0.859} & 36
& \textcolor{blue}{0.850} & 36
& \textcolor{ForestGreen}{0.908} & 24  \\

17 & RX\_RZ\_Multi\_CRX & DATA\_REUP\_PROB & AngleZ & 1.6190 & 0.955 % a6zvUHBWFnaKn4mr4a9SUC
& \textcolor{blue}{0.917} & 1193
& \textcolor{blue}{0.924} & 1192
& \textcolor{blue}{0.907} & 1150
& \textcolor{ForestGreen}{0.932} & 1100
& \textcolor{blue}{0.889} & 1413
& \textcolor{red}{0.654} & 555
& \textcolor{red}{0.653} & 559
& \textcolor{ForestGreen}{0.930} & 93\\

18 & RY\_RZ\_CNOT & DATA\_REUP\_PROB & AngleX & 1.4200 & 0.947 % XuuyQh7m9bpMz4oTVN8H7o
& \textcolor{ForestGreen}{0.941} & 162
& \textcolor{ForestGreen}{0.944} & 171
& \textcolor{ForestGreen}{0.941} & 134
& \textcolor{ForestGreen}{0.944} & 173
& \textcolor{ForestGreen}{0.937} & 239
& \textcolor{ForestGreen}{0.940} & 22
& \textcolor{ForestGreen}{0.940} & 22
& \textcolor{ForestGreen}{0.946} & 21 \\

19 & RXRZ\_CRZPool & DATA\_REUP\_PROB & AngleY & 2.5523 & 0.944 % ESVDh3ApsWXSda9tgPRv2b
& \textcolor{ForestGreen}{0.942} & 162
& \textcolor{ForestGreen}{0.944} & 171
& \textcolor{ForestGreen}{0.940} & 134
& \textcolor{ForestGreen}{0.943} & 173
& \textcolor{ForestGreen}{0.940} & 239
& \textcolor{ForestGreen}{0.942} & 22
& \textcolor{ForestGreen}{0.934} & 22
& \textcolor{ForestGreen}{0.941} & 21 \\

20 & H\_CZ\_RX & DATA\_REUP\_PROB & Amplitude & 1.5761 & 0.913 % dLSGnYdJaZQvBDSJFQwfQs
& \textcolor{red}{0.487} &  15354
& \textcolor{red}{0.456} & 15606
& \textcolor{red}{0.475} & 15414
& \textcolor{red}{0.519} & 15483
& \textcolor{red}{0.508} & 17811
& \textcolor{red}{0.482} & 6367
& \textcolor{red}{0.482} & 6365
& \textcolor{red}{0.500} & 5102\\

21 & RX\_RZ\_Multi\_CRZ & DATA\_REUP\_PROB & AngleY & 1.1158 & 0.927 % N2vraqCqvQC3XJLgJzJH32
& \textcolor{ForestGreen}{0.926} & 1124
& \textcolor{ForestGreen}{0.929} & 2083
& \textcolor{ForestGreen}{0.897} & 2063
& \textcolor{ForestGreen}{0.918} & 2172
& \textcolor{red}{0.821} & 2970
& \textcolor{red}{0.503} & 914
& \textcolor{red}{0.506} & 910
& \textcolor{ForestGreen}{0.923} & 137 \\

22 & QNN\_RY\_CNOT\_pool & VQC\_QNN\_EXP & AngleY & 1.6222 & 0.966 % MZWSKJxvdGKyZpCmc3Vpg6
& \textcolor{ForestGreen}{0.965} & 62
& \textcolor{ForestGreen}{0.963} & 62
& \textcolor{ForestGreen}{0.960} & 62
& \textcolor{ForestGreen}{0.962} & 62
& \textcolor{ForestGreen}{0.961} & 50
& \textcolor{ForestGreen}{0.952} & 26
& \textcolor{ForestGreen}{0.957} & 26
& \textcolor{ForestGreen}{0.960} & 33 \\

23 & RY\_CNOT & DATA\_REUP\_PROB & IQP & 0.0475 & 0.939 % cYeixNvqVZGWcccizUqLCk
& \textcolor{red}{0.504} & 2519
& \textcolor{red}{0.529} & 2519
& \textcolor{red}{0.519} & 2557
& \textcolor{red}{0.531} & 2580
& \textcolor{red}{0.491} & 3173
& \textcolor{red}{0.508} & 1005
& \textcolor{red}{0.500} & 989
& \textcolor{red}{0.524} & 405 \\

24 & RX\_RZ\_Multi\_CRZ & DATA\_REUP\_PROB & AngleY & 0.9210 & 0.949 %LBSuMzU5d74FKGKMdF6xKv 
& \textcolor{blue}{0.907} & 1100
& \textcolor{ForestGreen}{0.915} & 1102
& \textcolor{red}{0.841} & 1110
& \textcolor{ForestGreen}{0.937} & 1064 
& \textcolor{red}{0.796} & 1671
& \textcolor{red}{0.495} & 556
& \textcolor{red}{0.487} & 544
& \textcolor{ForestGreen}{0.951} & 93\\

25 & RXRZ\_CRZPool & DATA\_REUP\_PROB & IQP & 0.8288 & 0.926 % YgfFbCvKH2HaWBV2JBKc7X
& \textcolor{blue}{0.877} & 1359
& \textcolor{ForestGreen}{0.900} & 1445 
& \textcolor{blue}{0.861} & 1365
& \textcolor{ForestGreen}{0.907} & 1374 
& \textcolor{red}{0.796} & 1790
& \textcolor{red}{0.501} & 821
& \textcolor{red}{0.501} & 790
& \textcolor{ForestGreen}{0.906} & 227 \\

26 & RX\_RZ\_Multi\_CRZ & DATA\_REUP\_PROB & AngleY & 1.1158 & 0.927 % N2vraqCqvQC3XJLgJzJH32
& \textcolor{ForestGreen}{0.920} & 1167
& \textcolor{ForestGreen}{0.927} & 1063
& \textcolor{ForestGreen}{0.897} & 1113
& \textcolor{ForestGreen}{0.932} & 1122
& \textcolor{red}{0.811} & 1598
& \textcolor{red}{0.503} & 568
& \textcolor{red}{0.506} & 550
& \textcolor{ForestGreen}{0.923} & 93 \\

27 & RXRZ\_CRZPool & DATA\_REUP\_EXP & AngleY & 2.6991 & 0.920 % 8yvmDoufBUcjWWjdBoyzrf
& \textcolor{ForestGreen}{0.920} & 78
& \textcolor{ForestGreen}{0.922} & 74
& \textcolor{ForestGreen}{0.919} & 74
& \textcolor{ForestGreen}{0.922} & 86
& \textcolor{ForestGreen}{0.918} & 86
& \textcolor{ForestGreen}{0.915} & 47
& \textcolor{ForestGreen}{0.920} & 42
& \textcolor{ForestGreen}{0.918} & 13 \\

28 & RXRZ\_CNOT & DATA\_REUP\_EXP & AngleY & 0.0551 & 0.938 % dfa8wvnP3LwKqzYRfQsff7
& \textcolor{ForestGreen}{0.912} & 67
& \textcolor{ForestGreen}{0.923} & 67
& \textcolor{ForestGreen}{0.908} & 67
& \textcolor{ForestGreen}{0.920} & 67
& \textcolor{blue}{0.892} & 99
& \textcolor{red}{0.798} & 36
& \textcolor{red}{0.792} & 36
& \textcolor{ForestGreen}{0.904} & 21 \\

29 & RX\_RZ\_Multi\_CRX & DATA\_REUP\_EXP & AngleY & 0.0700 & 0.907 % M86Zi8iKCnpsB8UomYKyXJ
& \textcolor{red}{0.771} & 1115
& \textcolor{red}{0.774} & 1200
& \textcolor{red}{0.694} & 1180
& \textcolor{red}{0.819} & 1165
& \textcolor{red}{0.694} & 1517
& \textcolor{red}{0.491} & 524
& \textcolor{red}{0.474} & 547
& \textcolor{red}{0.838} & 93\\

30 & QNN\_RY\_CRX\_pool & VQC\_QNN\_EXP & AngleX & 1.0540 & 0.853 % 5zqHjiD6YZswBnLfre67kd
& \textcolor{ForestGreen}{0.857} & 58
& \textcolor{ForestGreen}{0.852} & 58
& \textcolor{ForestGreen}{0.860} & 58
& \textcolor{ForestGreen}{0.854} & 58
& \textcolor{ForestGreen}{0.856} & 62
& \textcolor{ForestGreen}{0.850} & 24
& \textcolor{ForestGreen}{0.854} & 24
& \textcolor{ForestGreen}{0.851} & 29  \\

31 & RXRZ\_CNOT & VQC\_QNN\_EXP & AngleY & 0.0930 & 0.913% eQLTRZefitqZG4fC9Saqq4
& \textcolor{ForestGreen}{0.929} & 67
& \textcolor{ForestGreen}{0.923} & 67
& \textcolor{ForestGreen}{0.916} & 67
& \textcolor{ForestGreen}{0.920} & 6767
& \textcolor{ForestGreen}{0.893} & 98
& \textcolor{red}{0.827} & 36
& \textcolor{red}{0.797} & 36
& \textcolor{ForestGreen}{0.922} & 24  \\

32 & RY\_CZ & VQC\_QNN\_EXP & Amplitude & 0.0830 & 0.899 % 5Bh5q4dMmXbpRgJDSJwzyG
& \textcolor{red}{0.644} & 1983
& \textcolor{red}{0.648} & 1979
& \textcolor{red}{0.538} & 1978
& \textcolor{red}{0.706} & 1964
& \textcolor{red}{0.526} & 2509
& \textcolor{red}{0.496} & 661
& \textcolor{red}{0.522} & 674
& \textcolor{red}{0.652} & 594 \\

33 & QNN\_RY\_CNOT\_pool & VQC\_QNN\_PROB & Amplitude & 0.2120 & 0.909 % YNSYjswNALtoRHPZuTPUMh
& \textcolor{red}{0.749} & 1614
& \textcolor{red}{0.818} & 1656
& \textcolor{red}{0.621} & 1521
& \textcolor{red}{0.756} & 1406
& \textcolor{red}{0.545} & 1817
& \textcolor{red}{0.470} & 661
& \textcolor{red}{0.482} & 632
& \textcolor{red}{0.790} & 535\\

34 & INV\_QFT\_U3\_MODEL & VQC\_QNN\_PROB & IQP & 0.1667 & 0.945 % TSAsu2jfWYSvD2Ai3YkpFa
& \textcolor{ForestGreen}{0.937} & 343
& \textcolor{ForestGreen}{0.940} & 342
& \textcolor{ForestGreen}{0.936} & 350
& \textcolor{ForestGreen}{0.940} & 399
& \textcolor{ForestGreen}{0.932} & 461
& \textcolor{ForestGreen}{0.708} & 189
& \textcolor{ForestGreen}{0.693} & 206
& \textcolor{ForestGreen}{0.937} & 50\\

35 & RY\_CNOT & VQC\_QNN\_PROB & IQP & 0.0380 & 0.933 % 2dTw986yWmXDn4acYToY8V
& \textcolor{blue}{0.872} & 378
& \textcolor{ForestGreen}{0.914} & 383
& \textcolor{red}{0.846} & 377
& \textcolor{ForestGreen}{0.904} & 382
& \textcolor{red}{0.717} & 506
& \textcolor{red}{0.500} & 176
& \textcolor{red}{0.517} & 142
& \textcolor{red}{0.850} & 72 \\

36 & RY\_CNOT & VQC\_QNN\_PROB & AngleY & 0.2970 & 0.907 % G6DDikD3QukvEUBQqeUq7X
& \textcolor{ForestGreen}{0.902} & 105
& \textcolor{ForestGreen}{0.902} & 106
& \textcolor{ForestGreen}{0.904} & 105
& \textcolor{ForestGreen}{0.904} & 105
& \textcolor{ForestGreen}{0.889} & 142
& \textcolor{red}{0.804} & 52
& \textcolor{red}{0.814} & 52
& \textcolor{ForestGreen}{0.899} & 36 \\

37 & RY\_CRX & VQC\_QNN\_PROB & IQP & 0.2750 & 0.906 % N5Gojw4Q23km4LPRUvZMF6
& \textcolor{ForestGreen}{0.897} & 355
& \textcolor{ForestGreen}{0.911} & 365
& \textcolor{ForestGreen}{0.891} & 355
& \textcolor{ForestGreen}{0.902} & 359
& \textcolor{ForestGreen}{0.897} & 442
& \textcolor{red}{0.728} & 174
& \textcolor{red}{0.736} & 175
& \textcolor{ForestGreen}{0.898} & 64 \\

38 & RXRZ\_CNOT & VQC\_QNN\_PROB & AngleX & 0.3167 & 0.888 % jWVWZ928SUwGWq8XWruXqo
& \textcolor{ForestGreen}{0.896} & 67
& \textcolor{ForestGreen}{0.890} & 67
& \textcolor{ForestGreen}{0.893} & 67
& \textcolor{ForestGreen}{0.891} & 67
& \textcolor{ForestGreen}{0.893} & 88
& \textcolor{ForestGreen}{0.875} & 36
& \textcolor{ForestGreen}{0.880} & 36
& \textcolor{ForestGreen}{0.887} & 24\\

39 & H\_CZ\_RX & VQC\_QNN\_PROB & AngleY & 0.2828 & 0.875 % MyGYPXPvYeZpdUUpmsbAnE
& \textcolor{ForestGreen}{0.874} & 54
& \textcolor{ForestGreen}{0.873} & 54
& \textcolor{ForestGreen}{0.869} & 54
& \textcolor{ForestGreen}{0.872} & 55
& \textcolor{ForestGreen}{0.873} & 144
& \textcolor{ForestGreen}{0.852} & 37 
& \textcolor{blue}{0.843} & 37
& \textcolor{ForestGreen}{0.869} & 37\\

40 & RY\_RZ\_CNOT & VQC\_QNN\_PROB & IQP & 0.0671 & 0.959 % f98uSxX5SBadxTzTea2Sqt
& \textcolor{red}{0.852} & 399
& \textcolor{blue}{0.881} & 398
& \textcolor{red}{0.841} & 379
& \textcolor{blue}{0.879} & 386
& \textcolor{red}{0.741} & 505
& \textcolor{red}{0.578} & 123
& \textcolor{red}{0.534} & 146
& \textcolor{red}{0.877} & 58\\

41 & RY\_RZ\_CNOT & DATA\_REUP\_EXP & IQP & 0.1131 & 0.935 % gJ4zypm9NS8Ma9TTLZiVG7
& \textcolor{red}{0.734} & 1279
& \textcolor{red}{0.795} & 1300
& \textcolor{red}{0.674} & 1281
& \textcolor{red}{0.838} & 1368
& \textcolor{red}{0.577} & 1873
& \textcolor{red}{0.490} & 721
& \textcolor{red}{0.490} & 718
& \textcolor{red}{0.858} & 197\\

42 & RXRZ\_CRZPool & DATA\_REUP\_PROB & AngleZ & 2.7730 & 0.930 % G6wwyieRx6zkuKnSWfgh56
& \textcolor{ForestGreen}{0.927} & 88
& \textcolor{ForestGreen}{0.928} & 73
& \textcolor{ForestGreen}{0.925} & 69
& \textcolor{ForestGreen}{0.928} & 73
& \textcolor{ForestGreen}{0.921} & 97
& \textcolor{ForestGreen}{0.927} & 24
& \textcolor{ForestGreen}{0.925} & 20
& \textcolor{ForestGreen}{0.928} & 13\\

43 & RXRZ\_CRZPool & DATA\_REUP\_PROB & IQP & 2.2790 & 0.923 % 72vuV3AwXWBViX87dzhWV7
& \textcolor{ForestGreen}{0.902} & 1413
& \textcolor{ForestGreen}{0.910} & 1415
& \textcolor{ForestGreen}{0.908} & 1382
& \textcolor{ForestGreen}{0.914} & 1361
& \textcolor{ForestGreen}{0.903} & 2057
& \textcolor{red}{0.552} & 845
& \textcolor{red}{0.560} & 776
& \textcolor{ForestGreen}{0.915} & 227 \\

44 & RXRZ\_CRZPool & DATA\_REUP\_PROB & IQP & 1.9780 & 0.892 % E784W3Krt4wggYHdWsmG6v
& \textcolor{ForestGreen}{0.902} & 1381
& \textcolor{ForestGreen}{0.905} & 1438
& \textcolor{ForestGreen}{0.899} & 1435
& \textcolor{ForestGreen}{0.905} & 1361
& \textcolor{ForestGreen}{0.892} & 1812
& \textcolor{red}{0.595} & 785
& \textcolor{red}{0.583} & 811
& \textcolor{ForestGreen}{0.906} & 227 \\

45 & RY\_RZ\_CNOT & DATA\_REUP\_EXP & AngleX & 0.2600 & 0.940 % 8XYHohM825ymKxoVL7XkVp
& \textcolor{ForestGreen}{0.924} & 180
& \textcolor{ForestGreen}{0.928} & 133
& \textcolor{blue}{0.915} & 177
& \textcolor{ForestGreen}{0.925} & 192
& \textcolor{blue}{0.899} & 182
& \textcolor{blue}{0.913} & 22
& \textcolor{blue}{0.915} & 22
& \textcolor{ForestGreen}{0.919} & 21 \\

46 & RXRZ\_CRZPool & DATA\_REUP\_EXP & AngleZ & 0.5350 & 0.892 % AknJjkcmz2EeapPDPSDn9E
& \textcolor{ForestGreen}{0.937} & 76
& \textcolor{ForestGreen}{0.941} & 69
& \textcolor{ForestGreen}{0.940} & 74
& \textcolor{ForestGreen}{0.941} & 78
& \textcolor{ForestGreen}{0.936} & 110
& \textcolor{ForestGreen}{0.934} & 28
& \textcolor{ForestGreen}{0.936} & 28
& \textcolor{ForestGreen}{0.934} & 13 \\

47 & QNN\_RY\_CRZ\_pool & VQC\_QNN\_EXP & AngleY & 1.5652 & 0.939 % emNpKD4Ty7mtErTQNWoVW7
& \textcolor{ForestGreen}{0.940} & 47
& \textcolor{ForestGreen}{0.941} & 47
& \textcolor{ForestGreen}{0.943} & 47
& \textcolor{ForestGreen}{0.943} & 47
& \textcolor{ForestGreen}{0.937} & 50
& \textcolor{ForestGreen}{0.933} & 18
& \textcolor{ForestGreen}{0.935} & 18
& \textcolor{ForestGreen}{0.936} & 29 \\

48 & RY\_CNOT & VQC\_QNN\_EXP & IQP & 0.0131 & 0.912 % XtJzfSiagthmQ88SKRfEZM
& \textcolor{red}{0.786} & 377
& \textcolor{red}{0.829} & 380
& \textcolor{red}{0.727} & 375
& \textcolor{red}{0.817} & 381
& \textcolor{red}{0.540} & 503
& \textcolor{red}{0.501} & 155
& \textcolor{red}{0.501} & 168
& \textcolor{red}{0.736} & 72\\

49 & RXRZ\_CNOT & DATA\_REUP\_PROB & AngleZ & 0.0318 & 0.902 % DQPPQeWfmxe6YLr4zeb3Rg
& \textcolor{blue}{0.859} & 67
& \textcolor{blue}{0.869} & 67
& \textcolor{blue}{0.830} & 67
& \textcolor{blue}{0.857} & 67
& \textcolor{blue}{0.834} & 98
& \textcolor{red}{0.708} & 36
& \textcolor{red}{0.668} & 36
& \textcolor{blue}{0.832} & 24\\

50 & RY\_CZ & DATA\_REUP\_PROB & AngleX & 0.0869 & 0.902 % EZCb44rHwhuxiwPPdD5sSc
& \textcolor{blue}{0.859} & 404
& \textcolor{blue}{0.869} & 403
& \textcolor{blue}{0.830} & 403
& \textcolor{blue}{0.857} & 398
& \textcolor{blue}{0.834} & 823
& \textcolor{blue}{0.708} & 92
& \textcolor{blue}{0.668} & 92
& \textcolor{blue}{0.819} & 92 \\

51 & RX\_RZ\_Multi\_CRZ & DATA\_REUP\_PROB & AngleZ & 0.5350 & 0.892 % 9vpxVxgcBv5VUg4N5QbTne
& \textcolor{red}{0.831} & 1192
& \textcolor{blue}{0.874} & 1125
& \textcolor{red}{0.771} & 1120
& \textcolor{blue}{0.877} & 1170
& \textcolor{red}{0.629} & 1613
& \textcolor{red}{0.500} & 571
& \textcolor{red}{0.433} & 541
& \textcolor{ForestGreen}{0.912} & 93\\
\bottomrule
\\

\end{tabularx}

\caption{Analysis of Variational Quantum Classification Models across Different Noisy Backends}

\label{tab:results}
\end{table}
\end{landscape}

\section{Discussion}\label{Discussion}

Our extensive experimentation across several variables such as datasets, model structures, devices, and ansatzes has proven that the results and interpretations presented in Section \ref{Results} are generalizable over a wide variety of Variational Quantum Classifiers.

The key observation from Table \ref{tab:results} is that what constitutes a noise-robust variational quantum classifier does not solely depend on depth or any single parameter but relies on other combinations of high average relative entropy amongst classes and low transpilation depth. This challenges the notion that a robust shallow variational circuit is one with a specific depth depending on the device's capability. 

We observed that several quantum circuits behaved differently on the same noisy devices, despite having similar transpilation depths. For instance, in the case of the noisy fake backend of the Sherbrooke device, we noted that a circuit with a depth of $2033$ showed a higher accuracy difference of $0.4$ between the simulation and the noisy device, while another circuit with a depth of $2236$ exhibited a much smaller accuracy difference of just $0.029$. This indicates that even with a greater depth, the accuracy difference can be minimal. The main distinction between these two circuits was their average relative entropy values. Many such observations can be derived from the data presented in Table \ref{tab:results}.

An additional observation from the graphs shown in Fig. \ref{fig:graph1}, Fig. \ref{fig:graph2}, and the log-DTSAE values of various devices in Section \ref{Results} is that each device has its own specific limits regarding the average relative entropy to transpilation depth ratio. This variation can be easily attributed to the differences in the noise characteristics of each of the noisy devices tested. We are currently exploring improved methods for calculating these values for noisy devices without requiring such extensive evaluations. 

The highest observed average relative entropy between classes was $8.571$, while the highest depth observed during experimentation was $20212$.

No model with low accuracy difference between simulated and noisy devices was observed for Amplitude encoding, as the transpilation depth for this method was notably high across all the noisy backends. The lowest observed transpilation depth for an Amplitude embedding model was $515$ for the IonQ Aria1, while IQM devices had a minimum value of $563$, with the lowest value for IBM devices reaching $1335$. These values are considerably high; for Amplitude embedding to be a viable option, the Average Relative entropy between classes needs to be extremely high. 

 Similar complexities for IQP embedding were observed, but weren't as severe as for Amplitude, as with certain lower depth bounds and a high average relative entropy between classes, the accuracy difference could be curbed. 

This brings about the problem with using these classical to quantum data encoding methods for practical Quantum Machine Learning implementation. Amplitude encoding can embed $n$ features into $log(n)$ qubits, but this ability is questioned due to the implementation complexity, along with its expressibility of classical data. Better methods for state preparation could widely help in alleviating this issue. 

There were other observations in the experiment that were associated with the transpilation procedure and the relative entropy during the data analysis. In a few observed models, the transpilation width, which is the number of qubits required to represent the circuit, increased due to the device structure being mapped to or the circuit design complexity. These lead to some values being skewed away from the norm, as showcased in Fig. \ref{fig:graph1} and \ref{fig:graph2}.

During the initial training phase, we also observed the same class average relative entropy, difference class average relative entropy, and their association with the model's accuracy. There isn't a straight correlation between the model's accuracy during simulation and the average entropy values. This leads to an important concern about how to train a model with a high average relative entropy between classes and high accuracy. Further work on this aspect is discussed in Section \ref{Future Work} and is currently under research.

As mentioned in Section \ref{Results}, we analyzed other metrics, such as the models' predictive probability distributions, and found no correlation with noisy device performance; however, these distributions substantially contribute to the models' accuracy. Skewed probability distribution in either of the classes leads to a model with less accuracy. 

Results from the real quantum device reflect a similar effect to the fake noisy backends. While a little more interference from noise is observed on the real quantum devices, the overall effect on the results was not magnified and does not contradict the relation between depth and relative entropy analyzed in this manuscript. More evaluations for real quantum device evaluations are in progress. 

Finally, based on the extensive experiments conducted on fake noisy backends, we propose some suggestions for the reproducibility of VQC models' results on noisy devices. The values for ideal transpilation depth and ideal average relative entropy between classes highly depend on the noisy device in question and the acceptable accuracy difference concerning the problem being reproduced. 

In Section \ref{Results}, we presented log-DTSAE graphs to showcase the dynamics of how these values influence each other, along with the accuracy difference, and it is clearly observed that these values are clearly device dependent and the noise channel associated with the model. But we suggest keeping the log-DTSAE ratio of models under 3.5 to have less deviation in accuracy. These values are bound to change with the progress in noise mitigation and betterment of gate fidelity. 

\section{Conclusion}\label{Conclusion}

%  work addresses two critical research gaps in Variational Quantum Classification (VQC) models: the varying impact of noise on model performance, and the uncertainty in defining an ideal depth for noise-robust shallow parameterized quantum circuits. Through extensive experiments across multiple VQC architectures, datasets, and noisy backends, we empirically establish a strong correlation between average relative entropy, transpilation depth, and the reproducibility of VQC results on noisy quantum hardware.

% Instead, a combination of average relative entropy between classes and circuit depth governs the reliability of VQCs under noise. To quantify this relationship, we introduce the log-DTSAE metric, which provides a classical, hardware-independent method to predict whether a VQC will reproduce simulated results on a noisy backend. 

%By providing a systematic approach to evaluate reproducibility, this work lays the foundation for scalable and reliable quantum machine learning procedures on NISQ devices, and offers practical guidance for designing noise-resilient VQC models.

In this article, we examine two critical research gaps in Variational Quantum Classification (VQC) models: the varying effects of noise on model performance in noisy quantum devices, and the uncertainty in defining the ideal depth for a noise-robust shallow parameterized quantum circuits.

Using empirical evidence, we establish the correlation between average relative entropy, transpilation depth, and the reliability of Variational Quantum Classifiers (VQCs) operating on a noisy backend. Our analysis is based on a large number of experiments conducted across various VQC architectures, datasets, and noisy quantum backends. 

We provide evidence that transpilation depth is not the only factor that determines noise resilience in shallow parameterized quantum circuits within VQCs. There is no single ideal depth value for any noisy device. Instead, a combination of the average relative entropy between classes and transpilation depth determines noise resilience in these models. 

To determine the reproducibility factor of a VQC model on noisy hardware, we propose a new metric: log-DTSAE (the logarithm of the depth divided by the square root of the average relative entropy). This method is also equally important to analyze the ideal transpilation depth in relation to the average relative entropy between classes in a noisy quantum backend

This research presents a classical method to evaluate the reproducibility of Variational Quantum Circuit (VQC) results on Noisy Intermediate-Scale Quantum (NISQ) hardware without the need for explicit executions on quantum devices. It lays the groundwork for scalable and reliable quantum machine learning (QML) procedures in noisy environments. Additionally, this article opens the door for further exploration of various other problems, which are discussed in the next section \ref{Future Work}.

\section{Future Work}\label{Future Work}
The observations made during the implementation of this study, along with the noted relationship between noise robustness and average relative entropy in variational circuits, suggest several promising directions for future research:

\begin{itemize}
    \item Develop computationally efficient methods to estimate log-DTSAE for new noisy quantum backends without requiring extensive empirical evaluations.
    
    \item Investigate warm-start strategies for initializing VQC models that promote higher average relative entropy between classes and improved robustness to noise.
    
    \item Explore training methodologies for Variational Quantum models that directly optimize average relative entropy on noisy quantum hardware.
    
    \item Analyze the scalability of the proposed relative entropy-based framework with increasing qubit counts and circuit complexity.

    \item Determining the ideal combination of multiple optimizer structures for faster minima approximation. 
    
    \item Study the impact of circuit compression techniques on models with low average relative entropy, particularly in improving their reproducibility on noisy backends.
\end{itemize}

\section{Acknowledgment}
We would like to express our special thanks to Valter Uotila for his valuable comments on the research and the creation of this manuscript. We also extend our gratitude to Ilmo Salmenperä, Frans Perkkola, Tung Bui Dang, and Leo Becker for their insightful discussions and guidance during the early phases of the research.

\section{Data and Code Repositories}\label{data}
 \begin{itemize}
     \item VQC-QNN Training and Testing Code and Data \cite{Shinde_PQCs_and_QNN_2026}:\url{https://github.com/AakashShindeHelsinki/VQC_QNN_QML_Model_Training_and_Benchmarking}
     \item Data-Reupload Training and Testing Code and Result Data \cite{Shinde_Data_Reupload_Probabilistic_2026}:\url{https://github.com/AakashShindeHelsinki/DataReupload_QML_Training_and_Benchmarking}
     \item QML model testing on noisy devices runs, transpilation depth generation, depth-entropy analysis, and data processing \cite{Shinde_QML_model_testing_2026}: \url{https://github.com/AakashShindeHelsinki/NoisyDev_Testing_and_Ent_Data_Processing}
 \end{itemize}
 
\section{Appendix}\label{Appendix}
    %write better descri[tion]
   Examples of all the circuit structures used in computation and experiment are available on the following GitHub repository: \url{https://github.com/AakashShindeHelsinki/Quantum_Circuit_Diagrams_for_RelEnt_analysis}

   %TEX DRWINGS TOO LARGE HIGH MEMORY REQUIREMENTS 
    %QNNS here
    
    %\input{Circuits/circuit_quantikz_export_batch_1}
    %\input{Circuits/circuit_quantikz_export_batch_13}
    %\input{Circuits/circuit_quantikz_export_batch_14}
    %\input{Circuits/circuit_quantikz_export_batch_15}
    %\input{Circuits/circuit_quantikz_export_batch_16}
    %\input{Circuits/circuit_quantikz_export_batch_17}
    %\input{Circuits/circuit_quantikz_export_batch_18}
    %\input{Circuits/circuit_quantikz_export_batch_19}
    %\input{Circuits/circuit_quantikz_export_batch_20}
    %\input{Circuits/circuit_quantikz_export_batch_21}
    %\input{Circuits/circuit_quantikz_export_batch_22}
    
    % VQCs here
    %\input{Circuits/circuit_quantikz_export_batch_2}
    %\input{Circuits/circuit_quantikz_export_batch_3}
    %\input{Circuits/circuit_quantikz_export_batch_4}
    %\input{Circuits/circuit_quantikz_export_batch_5}
    %\input{Circuits/circuit_quantikz_export_batch_6}
    %\input{Circuits/circuit_quantikz_export_batch_7}
    %\input{Circuits/circuit_quantikz_export_batch_8}
    %\input{Circuits/circuit_quantikz_export_batch_9}
    %\input{Circuits/circuit_quantikz_export_batch_10}
    %\input{Circuits/circuit_quantikz_export_batch_11}
    %\input{Circuits/circuit_quantikz_export_batch_12}
    
    %\section{VQC and Data-Reupload %Circuits}\label{data_vqc_circ}
    %DATA REUP here
    %\input{Circuits/Data_Reupload}

\bibliographystyle{ieeetr}
\bibliography{ref}

\end{document}